\documentclass[twocolumn]{aastex7}

\newcommand{\celmech}{\textsc \tt celmech}
\newcommand{\rebound}{\textsc \tt REBOUND}
\newcommand{\reboundx}{{\textsc \tt REBOUNDx}}
\usepackage{amsmath}

\begin{document}

\title{General Relativity Can Prevent a Runaway Greenhouse on Potentially Habitable Planets Orbiting White Dwarfs}

\author[orcid=0009-0001-2913-792X]{Eva Stafne}
\affiliation{University of Wisconsin-Madison, Department of Astronomy}
\email{stafne@wisc.edu}  

\author[orcid=0000-0002-7733-4522]{Juliette Becker} 
\affiliation{University of Wisconsin-Madison, Department of Astronomy}
\email{juliette.becker@wisc.edu}


\begin{abstract}
Planets orbiting in the habitable zones of white dwarfs have recently been proposed as promising targets for biosignature searches. However, since the white dwarf habitable zone resides at 0.01 - 0.1 AU, planets residing there are subject to tidal heating if they have any orbital eccentricity. Previous work \citep{BarnesHeller2013} identified nearby planetary companions as potential roadblocks to habitability of planets around white dwarfs, as such companions could induce secular oscillations in eccentricity for the potentially habitable planet, which could in turn heat a surface ocean and induce a runaway greenhouse for even very low values ($e \sim 10^{-4}$) of the eccentricity of the potentially habitable planet. In this work, we examine the potential for general relativistic orbital precession to protect habitable planets orbiting white dwarfs from such a runaway greenhouse, and demonstrate that for some system architectures, general relativity can be protective for planetary habitability. 
\end{abstract}



\section{Introduction} \label{intro}
White dwarfs, the remnants of low-mass stars, have been identified as potential hosts for habitable planets, both through theoretical studies demonstrating their feasibility \citep[e.g.,][]{Agol_2011, Fossati2012, Whyte_2024} and observational evidence of intact planets in orbits close to the habitable zone \citep[e.g.,][]{Vanderburg_2020, limbach2025thermalemissionconfirmationfrigid}.
Although no intact habitable planet candidates have yet been confirmed around white dwarfs (likely due to observational challenges such as low transit probabilities and short transit durations), this region of parameter space is especially compelling for biosignature studies \citep{Loeb_2013}. If an Earth-like planet were discovered within the habitable zone of a white dwarf, the combination of deep transit signals and the favorable spectral characteristics of white dwarfs could enable the detection of atmospheric biosignatures using JWST NIRSpec transmission spectroscopy \citep{Kaltenegger_2020} or via IR excess with MIRI/MRS \citep{Limbach_2022_methods}.

There are significant challenges to placing planets in the habitable zone of a white dwarf: the habitable zone of a white dwarf resides at 0.01-0.1 AU, well interior to the stellar remnant's inferred past radius when it was on the red giant branch, and the size of the habitable zone shrinks as a white dwarf cools \citep{Agol_2011}. 
The presence of objects at radii where they should have been engulfed during the red giant phase \citep[e.g.,][]{Vanderburg2015, Vanderbosch2020, Limbach2024} suggests that planet formation and migration remain active processes after the main sequence. 

Planets orbiting white dwarfs could either have existed on the main sequence and survived stellar evolution \citep{Debes2002,  Veras2013, Adams2013, Veras2016}, or have formed after the star started to evolve off of the main sequence \citep{Perets2010}, possibly from the debris from tidally disintegrated planets \citep{Veras2020}, a disrupted stellar companion \citep{Chamandy2025}, or common envelope ejecta \citep{Schleicher2014}. The formation of planets after the host star has left the main sequence has been observed to create multi-planet systems around neutron stars \citep{Wolszczan1992} and is expected to do so around white dwarfs as well \citep{Bear2015, vanLieshout2018}. 

One challenge to the habitability of planets in the orbiting white dwarfs is the possibility of a runaway greenhouse effect \citep{Kasting1988}, a climate state in which a planet, despite having an apparently habitable effective temperature, becomes excessively insulated by greenhouse gases such as water vapor or CO$_{2}$, driving surface temperatures beyond habitable limits.
\cite{BarnesHeller2013} first identified this issue, showing that in multi-planet systems around white dwarfs, a planet located in the habitable zone can have its orbital eccentricity excited by interactions with neighboring planets. This non-zero orbital eccentricity leads to tidal heating for planets sufficiently close to the central body \citep[e.g.,][]{1979Sci...203..892P, Driscoll2015, Barr2018, Seligman2024}, in ways that can significantly affect planetary habitability \citep{ra00500s, Barnes2009}. Planets in a white dwarf's habitable zone are particularly susceptible to the effects of tidal heating due to their short-period orbits \citep{BarnesHeller2013, Becker2023}. While a lone planet could circularize its orbit and avoid this issue \citep[e.g.,][]{Veras2019}, in a multi-planet system, eccentricities are continually forced. As a result, the persistent tidal heating can drive the planet’s climate into a runaway greenhouse state, even at relatively low orbital eccentricities \citep[$e \leq 10^{-4} - 10^{-6}$;][]{BarnesHeller2013}.


However, since the habitable zone is so close to the white dwarf, planets residing there will be subject to significant apsidal precession due to general relativity \citep[GR;][]{Einstein1915}. The orbital precession induced by GR could potentially perturb the dynamics of a multi-planet system, altering the amplitude of secular eccentricity cycles, and subsequently the degree of tidal heating that the planet will experience. 
In this paper, we evaluate how the effects of orbital precession due to GR affects the onset of a runaway greenhouse atmosphere. 
In Section \ref{sec:theory}, we present a secular framework to model the eccentricity evolution of the inner planet under the influence of both general relativistic precession and planet-induced libration.
In Section \ref{sec:results}, we present the results of our parameter sweep study, conducted using the open source code \celmech.
In Section \ref{sec:discussion}, we discuss the significance and implications of our results. 
Finally, in Section \ref{sec:conclusion}, we conclude with a summary of our results. 

\section{Analytic Framework for GR-Stabilized Orbits} \label{sec:theory}
In this section, we explore how GR might affect planet-planet interactions for planets orbiting white dwarfs. 
Planet-planet interactions naturally lead to secular oscillations in orbital eccentricity as planets exchange angular momentum \citep{MD99, Batygin2013}, which, in turn, will result in potentially substantial tidal heating \citep{Hut1981} for close-in planets orbiting white dwarfs  \citep{BarnesHeller2013, Becker2023}. 
As demonstrated by \citet{BarnesHeller2013}, this process can trigger a runaway greenhouse effect, rendering an otherwise potentially habitable planet uninhabitable.
In the past, it has been demonstrated that GR can significantly alter multi-planet dynamics \citep{Adams2006, Faridani2022, Iorio2023}, potentially even suppressing mechanisms like Lidov-Kozai resonance \citep{Volpi2024}. 
In this section, we present motivating analytic equations to model where in parameter space multi-planet interactions capable of causing a runaway greenhouse would occur, and whether the inclusion of GR effects could prevent the runaway greenhouse from being triggered.

\begin{figure}
    \centering
    \includegraphics[width=0.95\linewidth]{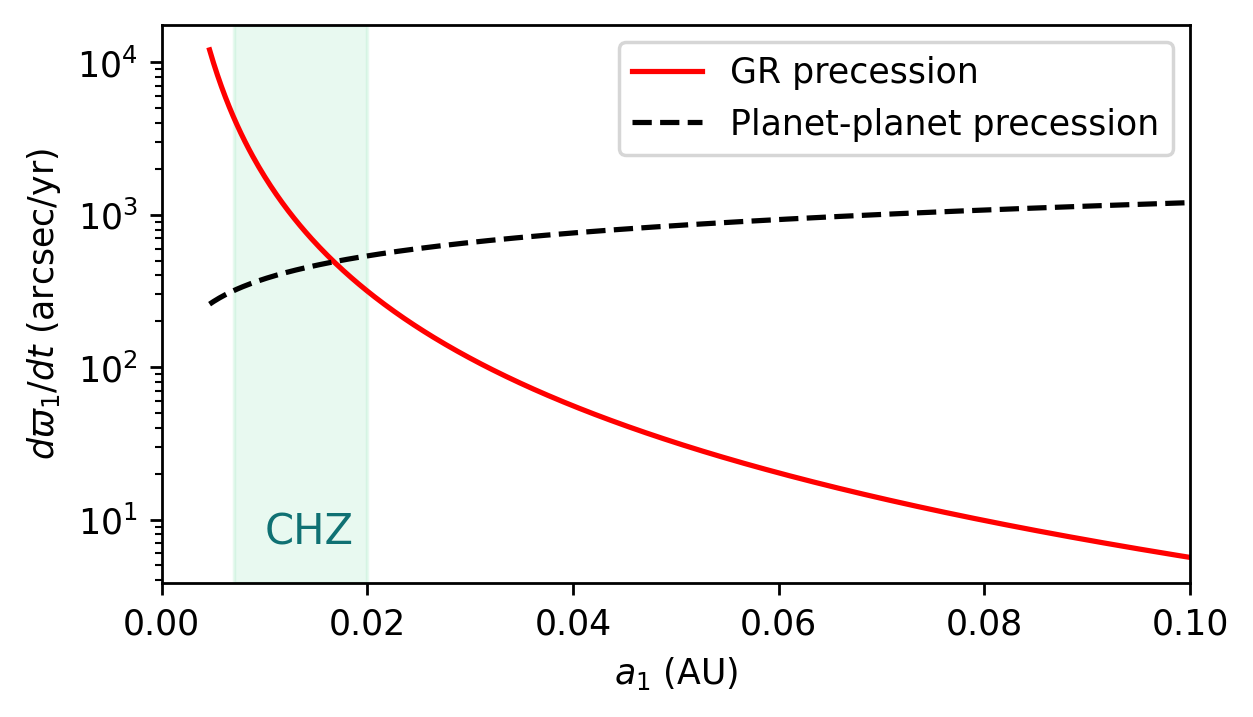}
    \caption{The plot shows the precession rate for an inner planet $a_1$ perturbed by an exterior companion planet with mass $m_2 = 10 M_{\oplus}$ at a orbital distance of $a_2 = 0.1$ AU. The red curve represents the precession rate due to GR, while the dashed black curve represents the precession rate due to perturbations from the exterior companion. The shaded green region highlights the range corresponding to the Continuous Habitable Zone (CHZ) identified in \citet{Agol_2011}.}
    \label{fig:CHZ_precession}
\end{figure}

\subsection{Precession Rates: General Relativity vs. Planetary Interactions}
To evaluate approximately the relative importance of GR and planet-planet interactions, we can compare the relative precession rates induced on a planet's argument of periastron from each source. 
The magnitude of these two contributions can be expressed analytically. The effect of GR on orbital precession for a planet orbiting at a semi-major axis $a_{1}$ is \citep[e.g.,][]{2019PhR...805....1B, Faridani2025}:
\begin{equation}
\left(\frac{d\varpi_1}{dt}\right)_{\mathrm{GR}} = 3\sqrt{\frac{GM\star}{a_1^3}} \cdot \frac{GM_\star}{a_1c^2}
\label{eq:GR}
\end{equation}
where subscripts 1 and 2 denote the inner and outer planets respectively, $\varpi$ is a planet's longitude of pericenter, $G$ is the gravitational constant, $M_\star$ is the mass of the central star, $a$ denotes the planetary semi-major axis, $c$ is the speed of light, and $e$ is the plaentary orbital eccentricity. 
The time rate of change of orbital precession $d\varpi/dt$ induced by planet-planet interactions due to an exterior companion planet is given by \citep[e.g.,][]{MD99, Anderson2017, 2019PhR...805....1B}:
\begin{equation}
    \left(\frac{d\varpi_1}{dt}\right)_{\mathrm{pl-pl}} = 
    \frac{3 n_1}{4} \frac{m_2}{M_*} \, \Big(\frac{a_1}{a_2}\Big)^2 
    \label{eq:OP}
\end{equation}
where $m$ denotes a planet's mass.

In Figure \ref{fig:CHZ_precession}, we show the relative precession rates induced by GR and planet-planet perturbations caused by an exterior companion (with $m_2 = 10 M_{\oplus}$, $a_2 = 0.1$ AU) for an interior planet at a range of semi-major axis $a_1$. We also show as a shaded region the Continuous Habitable Zone (CHZ), defined in \cite{Agol_2011} as the region around a white dwarf where a planet will remain habitable for 3 Gyr or more. 

While the specifics of which precession rate dominates inside the CHZ will depend sensitively on specific planet parameters, the important takeaway from this plot is that due to how close the white dwarf CHZ is to its host star, GR precession can dominate over planet-planet interactions for sufficiently close-in inner planets. 
The next question we must address is how GR will affect the amplitude of eccentricity oscillations, which we will do using the Laplace-Lagrange analytical framework.

\subsection{The Laplace-Lagrange Framework}
\label{sec:LLtheory}
In the secular Laplace-Lagrange framework, a planet's Keplerian motion through its orbit is ignored, with the orbit instead being mass-averaged. This approximation works well for non-resonant systems where the dynamical evolution does not depend on the mean longitudes of the planets. 
In this formulation, the disturbing function $\mathcal{R}_1$ for the motion of a planet with a mass $m_{1}$ and semi-major axis $a_1$ perturbed by an exterior companion (with mass $m_{2}$ and semi-major axis $a_2$) can be written as \citep{MD99}:
\begin{equation}
\begin{split}
    \mathcal{R}_1 = n_1 a_1^2 \bigg[ \frac{1}{2} A_{11} e_1^2 &+  A_{12} e_1 e_2 \cos(\varpi_1 - \varpi_2) \\
    & + A_{21} e_1 e_2 \cos(\varpi_2 - \varpi_1) \bigg] .
\end{split}
\end{equation}
where the $A_{11}$ matrix element has been adapted to include the effect of GR precession, Equation \ref{eq:GR}:
\begin{equation}
A_{11} = n_1 \left[ \frac{1}{4} \frac{m_2}{M_* + m_1} \, \Big(\frac{a_1}{a_2}\Big)^2 \, b_{3/2}^{(1)}(a_1/a_2) + \frac{3GM_*}{c^2 a_1} \right]
\end{equation}
and as in \citet{MD99},
\begin{align}
A_{12} = -n_1 \, \frac{1}{4} \, \frac{m_2}{M_* + m_1} \, \Big(\frac{a_1}{a_2}\Big)^2 \, b_{3/2}^{(2)}(a_1/a_2) \\
A_{21} = -\,n_{2}\,\frac{1}{4}\,\frac{m_{1}}{M_\ast + m_{2}}\,\Big(\frac{a_1}{a_2}\Big) b^{(2)}_{3/2}\!\left(a_1/a_2\right),
\end{align}
where $b_{3/2}$ denotes the Laplace coefficient \citep{MD99}, and $b^{(2)}_{3/2}\!\left(a_1/a_2\right) \approx 3 a_1/a_2$ if $a_1 \ll a_2$.


Using this framework, \citet{Adams2006} gives a criterion (in their Equation 13) dividing the parameter spaces where GR will amplify or suppress eccentricity oscillations due to a nearby companion planet:  
\begin{equation}
\xi = \frac{m_2}{m_1}\left[ \left( \frac{a_2}{a_1} \right)^{1/2} + \frac{4 G M_\ast^2}{c^2 a_1 m_2} \left( \frac{a_2}{a_1} \right)^{7/2}\right],
\label{eq:xi}
\end{equation}
where if $\xi<1$, GR will amplify eccentricity oscillation amplitudes due to an exterior companion planet, and when $\xi>1$, GR will suppress them. 

\begin{figure}[ht]
\centering
\includegraphics[width=0.47 \textwidth]{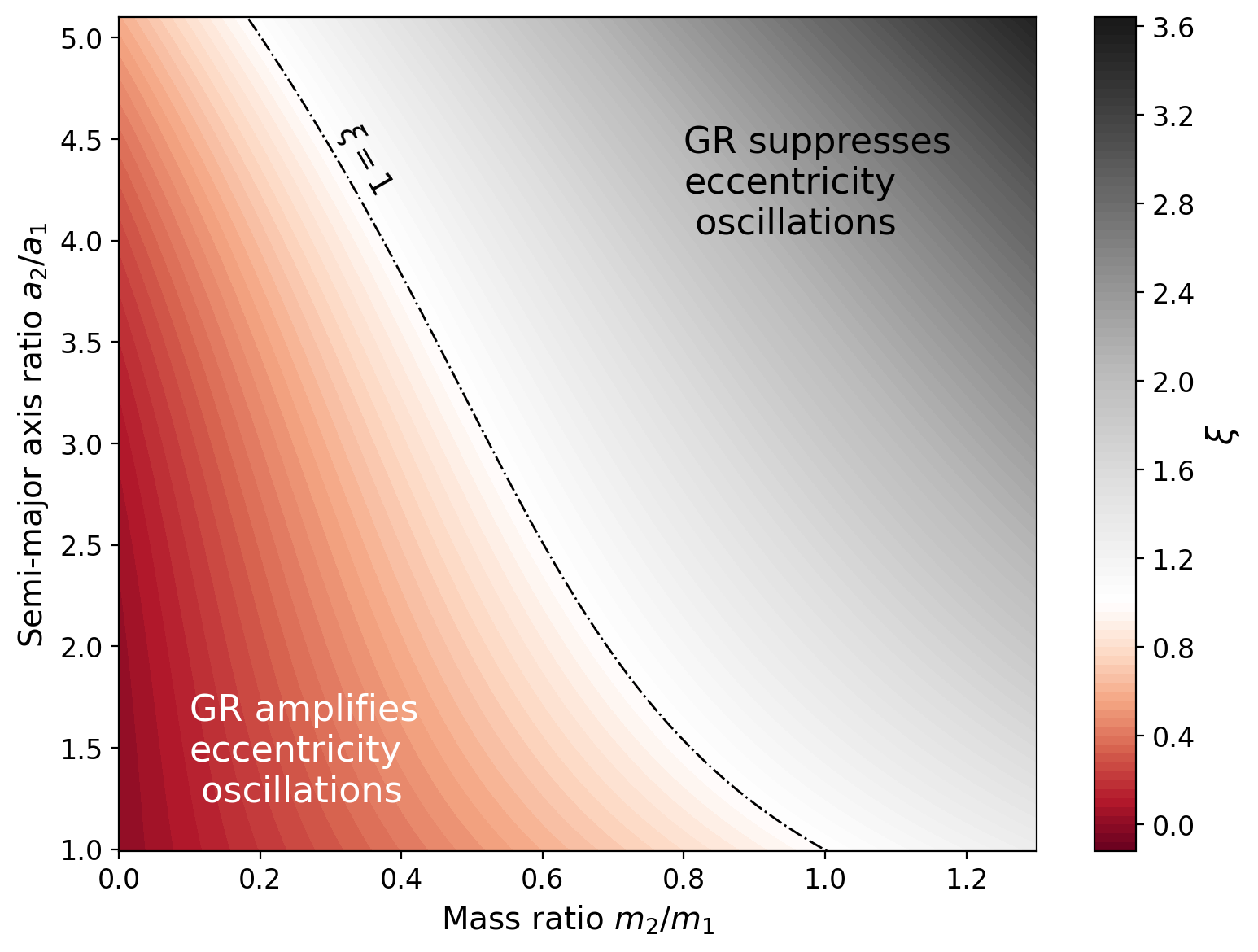}
\caption{Parameter space showing the effect of GR on eccentricity oscillations due to an exterior planetary companion for a variety of combinations of the mass ratio $m_2/m_1$ and semi-major axis ratio $a_2/a_1$. The interior planet was assumed to be a 1 $M_{\oplus}$ planet at 0.01 AU. When $\xi < 1$ (bottom-left, red), GR amplifies eccentricity oscillations; when $\xi > 1$ (top-right, black), GR suppresses them. The dashed line marks the transition as derived in \citet{Adams2006}.
\label{fig:xi_boundary}}
\end{figure}

In this work, we are interested in the region of parameter space where $\xi > 1$, as in this region, GR will act to counteract eccentricity oscillations induced by a companion, potentially preventing the tidal-heating-induced runaway greenhouse described in \citet{BarnesHeller2013} to be a challenge to habitability. 
We note that the region of the plot where $\xi < 1$ may act to increase the amplitude of eccentricity oscillations, and more readily create a runaway greenhouse.


\section{Results} \label{sec:results}
As discussed in the previous section, in systems with mass ratios $m_2>m_1$ the effect of GR precession will be to suppress eccentricity oscillations. 
However, the degree to which eccentricity oscillations are suppressed will depend on the particular system parameters under consideration. 
\cite{BarnesHeller2013} identified the threshold for the onset of a runaway greenhouse effect on a planet in the habitable zone of a white dwarf, finding that this threshold occurs at an orbital eccentricity between $10^{-4}$ and $10^{-6}$. The specific eccentricity depends on the age of the white dwarf and the planet's orbital radius (older white dwarfs with closer planets corresponding to the $e\simeq10^{-6}$ threshold).

In this section, we will consider a potentially habitable planet around a white dwarf with a semi-major axis $a_1 = 0.01$ AU and a mass $m_1 = 1 M_{\oplus}$, perturbed by an exterior planet with varying parameters
, and determine in what parameter space the effect of GR precession is sufficient to protect the inner planet from a runaway greenhouse.

\subsection{Simulation Set-Up} \label{sec:methods}

To solve the equations of motion using secular Laplace-Lagrange theory as outlined in Section \ref{sec:LLtheory}, we use the open-source secular integration package \celmech\ \citep{Hadden_2022}. 
This secular model allows us to model the evolution of an inner planet's eccentricity due to effects of both an outer planet and GR precession. For a brief discussion of our use of the secular model as opposed to a full N-body simulation, see Appendix \ref{app:nbody}.
In the \celmech\ Hamiltonian, we include secular terms up to second order in eccentricity.
For each set of initial conditions, we run two integrations, 
one with and one without GR precession, to quantify how it affects the parameter space where a runaway greenhouse occurs for multiple system geometries. 

In all of our integrations, the central body is a 0.5 $M_{\odot}$ white dwarf, and the inner planet is initialized with mass $m_1 = 1 M_{\oplus}$ at a semi-major axis of $a_1 = 0.01$ AU, within the expected habitable zone for a cooling white dwarf \citep{Agol_2011}. 
The initial eccentricity of the inner planet is set to $e_1 = 0$, under the assumption that any primordial free eccentricity would have been damped by tidal forces over time. 
As a result, any inner planet eccentricity induced during the simulation is an effect of the outer planet's secular influence, potentially driving the inner planet into a runaway greenhouse state.

\begin{figure}[ht]
\includegraphics[width=\columnwidth]{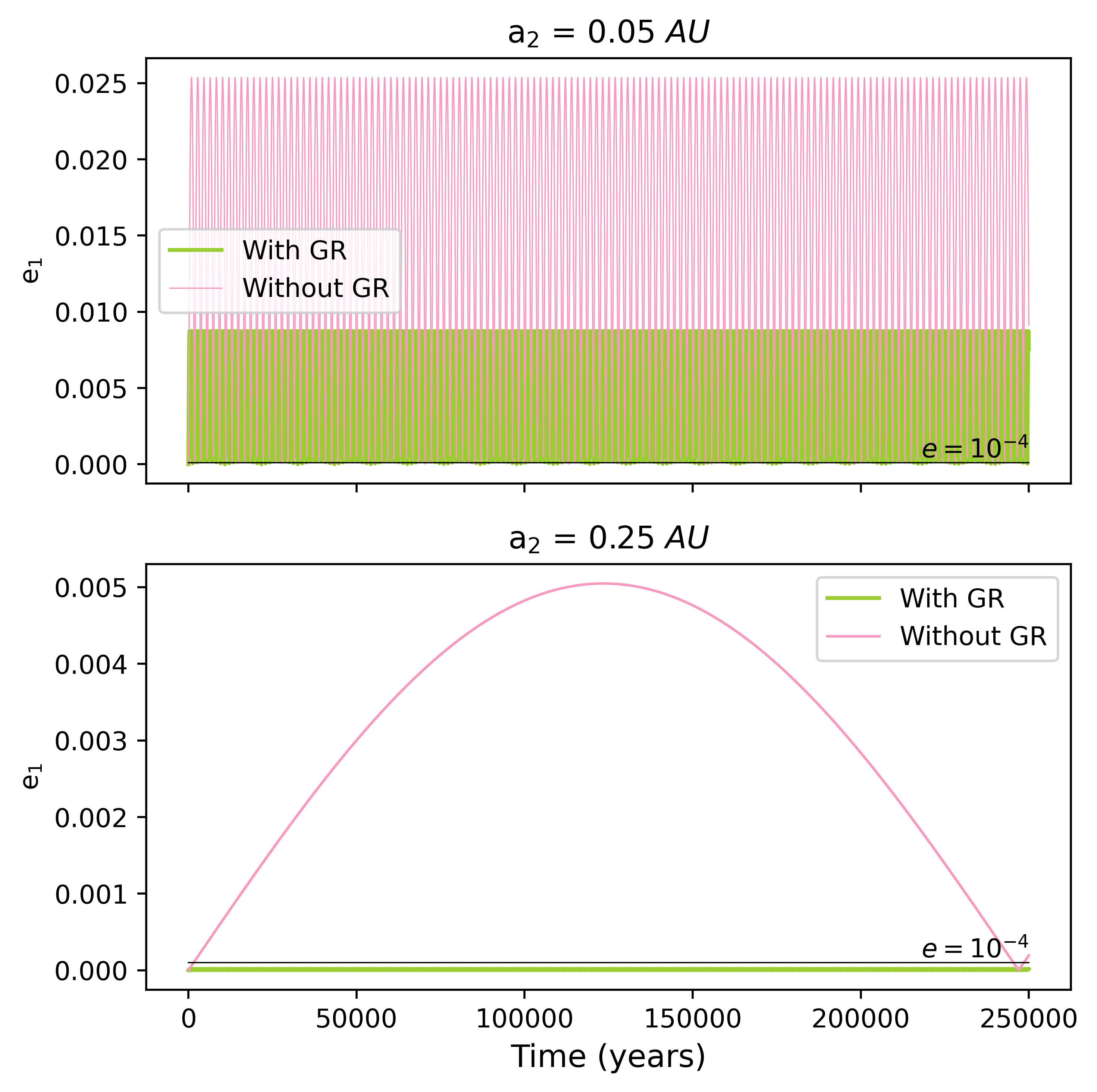}
\caption{Example integrations of the inner planet's eccentricity $e_1$ for two values of the outer planet's semi-major axis $a_2$, with all other system parameters identical. In each panel, the pink curve shows the solution without GR and the green curve includes GR. The top panel corresponds to $a_2 = 0.05\,\mathrm{AU}$ and the bottom to $a_2 = 0.25\,\mathrm{AU}$. The outer planet mass is set at a value of 20 $M_\oplus$. The eccentricity oscillations are larger for the closer outer planet, and in both cases the inclusion of GR suppresses their amplitude.}
\label{fig:astack}
\end{figure}

\begin{figure}[!h]
\includegraphics[width=\columnwidth]{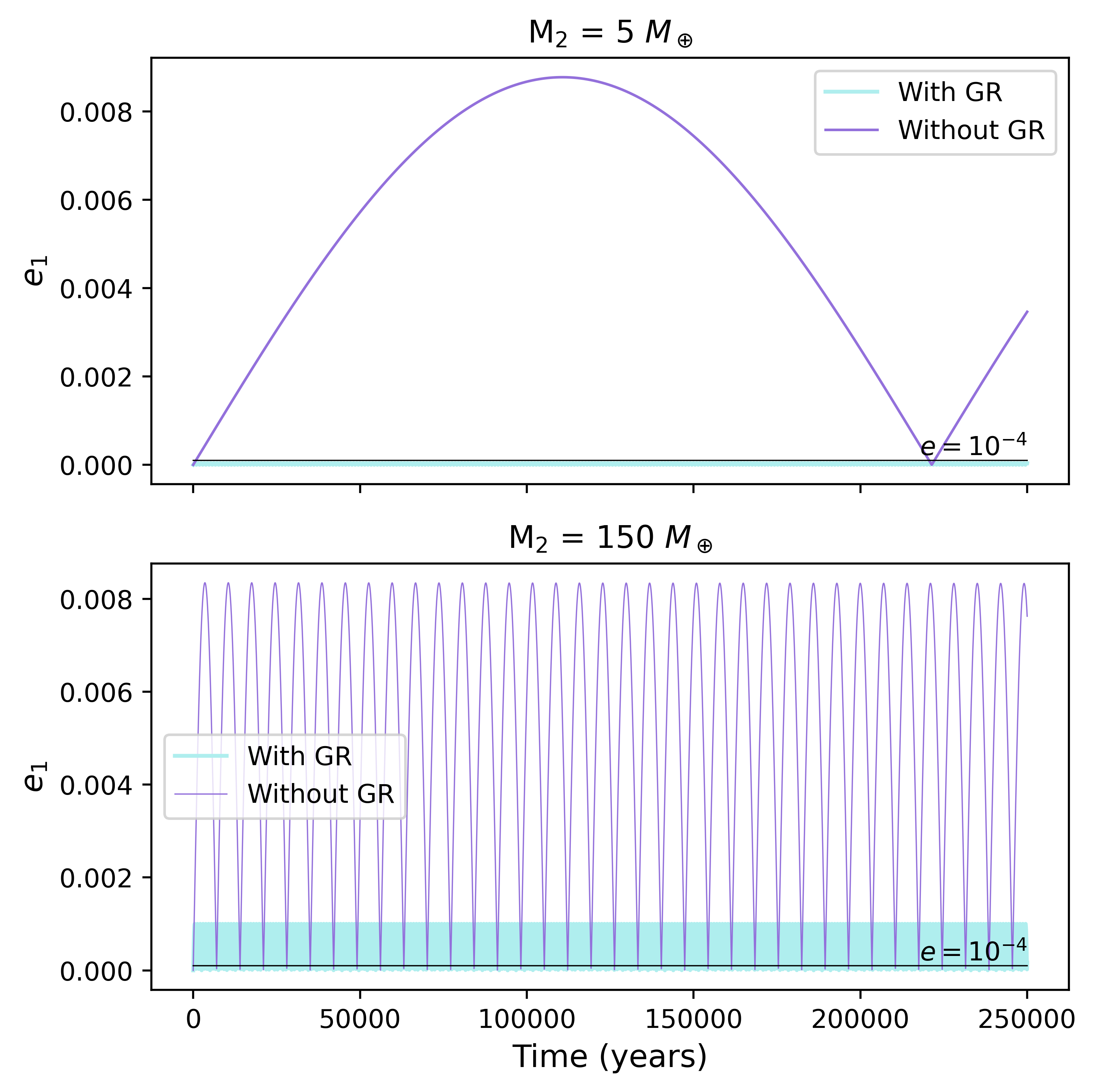}
\caption{Example integrations of the inner planet’s eccentricity $e_1$ for two values of the outer planet’s mass $m_2$ ({top panel:} $m_2 = 5\,M_\oplus$; {bottom panel:} $m_2 = 150\,M_\oplus$), with all other system parameters identical and the outer planet semi-major axis set at a value of 0.15 AU. Larger outer planet masses lead to higher-frequency eccentricity oscillations. In both cases, the inclusion of GR strongly suppresses the oscillation amplitude, more strongly for smaller masses of the outer planet. }
\label{fig:mstack}
\end{figure}

\begin{figure*}[!h]
\includegraphics[width=0.95\textwidth]{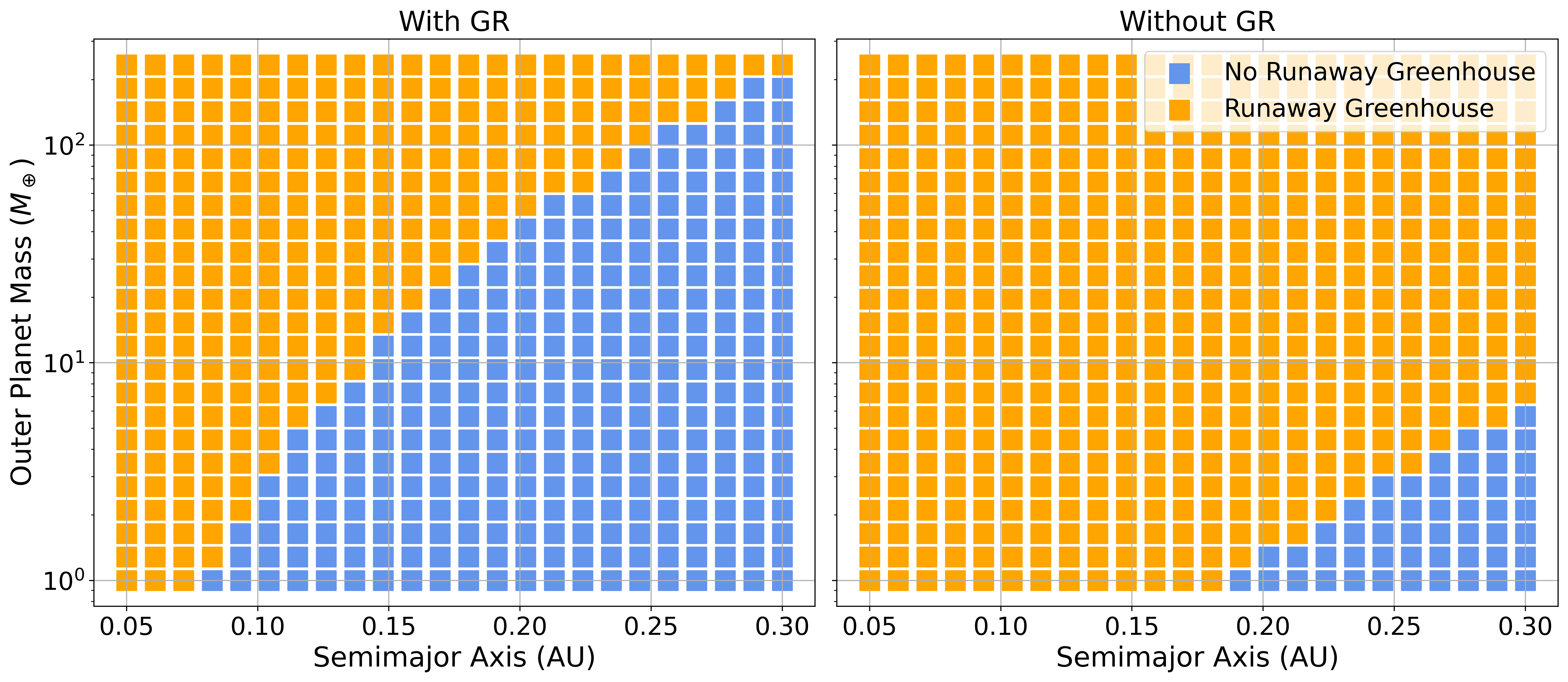}
\caption{Parameter sweep over outer planet mass ($m_2$) and semi-major axis ($a_2$), testing whether a $1\,M_\oplus$ inner planet at $a_1=0.01$ AU orbiting a $0.5\,M_\odot$ white dwarf undergoes a tidal-driven runaway greenhouse, if the runaway greenhouse will onset at the level of tidal heating caused by a orbital eccentricity of $e\approx10^{-4}$. \emph{Left panel:} A parameter sweep which includes GR apsidal precession. \emph{Right panel:} an identical parameter sweep but without GR. GR enlarges the stable region (where the inner potentially habitable planet avoids the runaway greenhouse) by suppressing secularly forced eccentricity, whereas without GR nearly all configurations with $a_2 \lesssim 0.18$ AU trigger a runaway greenhouse. All runs assume $e_2=0.05$ and are integrated for $3\times10^{5}$ yr.}
\label{fig:Csweep}
\end{figure*}

In Figure \ref{fig:astack}, we show two example solutions for the evolution of the inner planet's eccentricity $e_1$ for two values of $a_2$, for otherwise identical system parameters. The outer planet has a mass of 20 $M_\oplus$ and eccentricity of 0.05.
On each panel, the pink curve shows the solution without the inclusion of GR, and the green curve shows the solution with GR included. 
The top panel corresponds to an outer companion planet at $a_2 = 0.05$ AU, and the bottom panel to $a_2 = 0.25$ AU. 
The amplitude of the eccentricity oscillation is larger for the nearer outer planet, as expected. 
For both values of $a_2$, the inclusion of GR reduces the amplitude of the eccentricity oscillations as compared to the solution without GR, as expected.
When the outer planet is farther away ($a_2 =0.25$ AU), general relativity prevents a runaway greenhouse by keeping eccentricity oscillations low, but when it is closer ($a_2 = 0.05$ AU), strong interactions drive the inner planet to uninhabitable conditions regardless of GR's effect.

In Figure \ref{fig:mstack}, we show an analogous plot to Figure \ref{fig:astack}, but with the mass of the exterior planet set to be either $m_2 = 5 M_{\oplus}$ or $m_2 = 150 M_{\oplus}$ but with otherwise identical parameters. The outer planet is set to have a semi-major axis $a_2 = 0.15$ AU.  
In this plot, the integrations without GR exhibit larger eccentricity oscillations compared to those with GR included.
Both of the examples shown in Figure \ref{fig:astack} and Figure \ref{fig:mstack} demonstrate the way that GR precession can decrease the amplitude of planet-planet induced eccentricity oscillations. 

\subsection{Simulation Results} 


Using the set-up described in the previous sub-section, we conduct a two-dimensional parameter sweep across outer planet mass ($m_2$) and semi-major axis ($a_2$), for a total of 576 integrations. We vary the mass of the outer planet $m_2$ between 1 to 300 $M_{\oplus}$, sampled logarithmically with 24 values, while semi-major axis $a_2$ is sampled linearly from 0.05 to 0.30 AU, also with 24 values. 
All outer planets are initialized at an eccentricity of 0.05. Simulations are run for $3 \times 10^5$ years in order to capture full secular cycles for all sampled configurations.

Following the results of \citet{BarnesHeller2013}, we assume that for our Earth-like inner planet at $a_1= 0.01$ AU, a runaway greenhouse will onset if the planetary eccentricity reaches $e = 10^{-4}$. If that occurs, then the planet will no longer be habitable. 
Figure~\ref{fig:Csweep} shows the maximum eccentricity $e_1$ of the inner planet for each of our simulated parameter combinations, both including (left panel) and excluding (right panel) GR.  
As shown in the right panel of Figure~\ref{fig:Csweep}, without accounting for the effects of general relativity, the vast majority of the parameter space for a companion, including all orbits interior to $0.18$ AU, would be expected to induce eccentricity variations large enough to trigger a runaway greenhouse effect, resulting in uninhabitable conditions for the inner planet.
However, once GR is added into the model (left panel of Figure~\ref{fig:Csweep}), the range of outer planet parameters where the inner, potentially habitable planet avoids the runaway greenhouse becomes significantly larger.

We find that outer planet masses above 250 $M_{\oplus}$ universally exceed this threshold, rendering the inner planet uninhabitable even when general relativity is included. These results are consistent with the predictions of \citet{BarnesHeller2013}, in that even with general relativistic precession, sufficiently massive outer planets can still drive the inner planet’s eccentricity above the runaway greenhouse threshold. As expected from secular theory, general relativity is more effective in stabilizing inner planets in systems with less massive or more distant outer planets. 



\begin{figure}[h]
\plotone{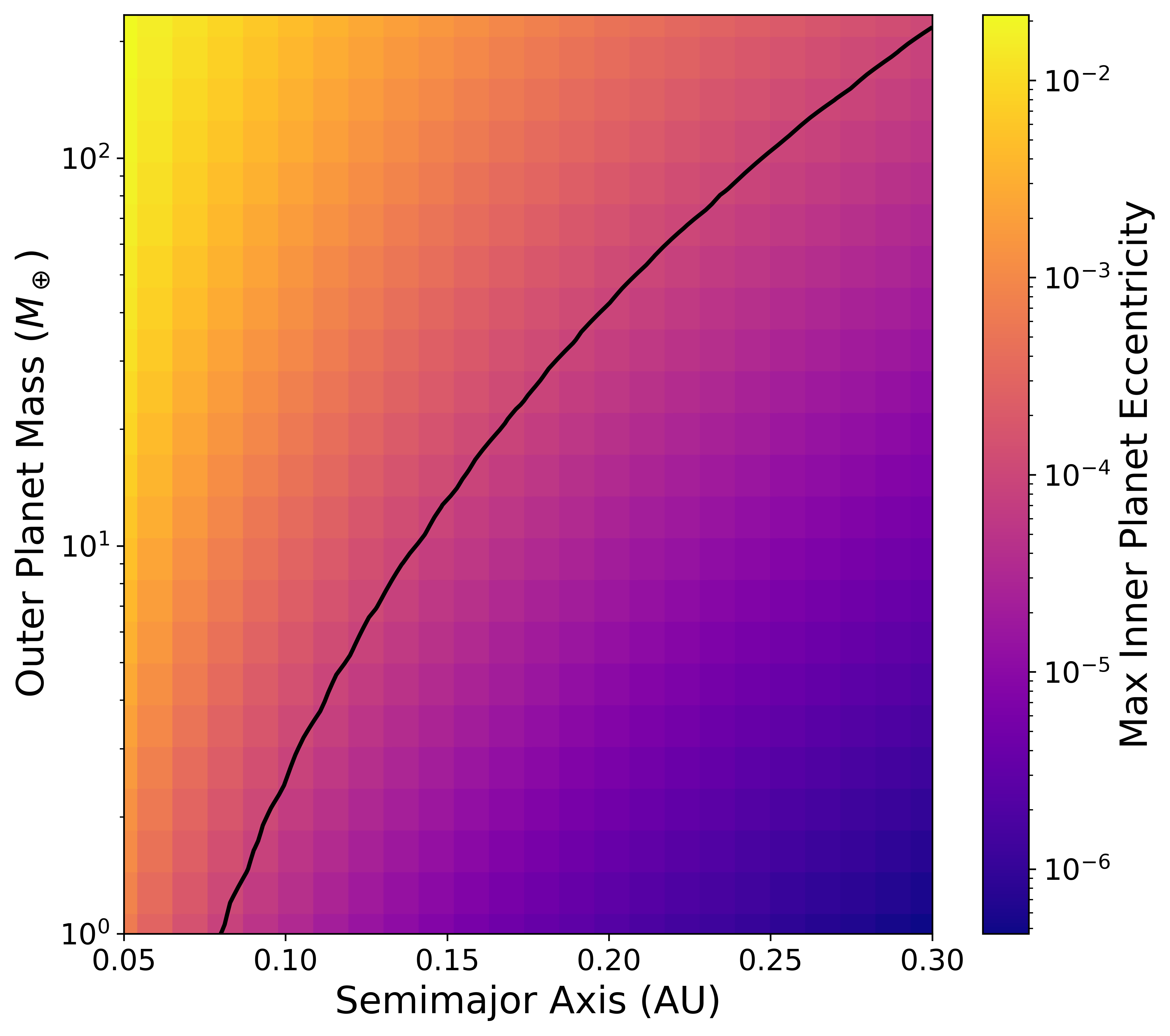}
\caption{Maximum eccentricities of an inner planet with varying outer planet parameters. Contour shows runaway greenhouse effect cutoff of e = 10$^{-4}$}
\label{fig:colormap}
\end{figure}

In Figure \ref{fig:colormap}, we present similar parameter sweep to that of Figure \ref{fig:Csweep}, but only shows what happens if we include general relativity terms in our simulations. 
This plot illustrates the maximum computed eccentricity of the inner planet for a range representative of plausible system geometries for white dwarf planetary systems. The contour at $e = 10^{-4}$ indicates the threshold beyond which a runaway greenhouse effect is expected. Potentially habitable planets in systems containing an exterior companion with parameters to the left of this line will likely not be habitable due to the onset of a runaway greenhouse. However, parameter combinations to the right of this line will allow the potentially habitable planet to avoid the runaway greenhouse. 

Together, these results reveal the importance of general relativity in stabilizing habitable zone planets in white dwarf systems. Without general relativity, the predicted habitability of an inner planet is significantly reduced. Even so, the parameter space that permits habitability shrinks further for more massive or more tightly packed planetary systems, consistent with expectations from secular theory.

\section{Discussion} \label{sec:discussion}
In this paper, we considered the result of \citet{BarnesHeller2013}, which found that planets orbiting in the habitable zone of a white dwarf which are perturbed by nearby planetary companions may have a runaway greenhouse atmosphere caused by tidal heating due to planet-planet interactions. In this work, we evaluated whether general relativity (GR) affects this mechanism. 
The results of our secular parameter sweep demonstrate that there is a significant parameter space of outer planet configurations where a planet in the habitable zone of a white dwarf would be uninhabitable due to the onset of a runaway greenhouse, if the effects of GR are not modeled in the system. 
However, if GR precession is included in the model, those configurations avoid the onset of a runaway greenhouse, potentially preserving the habitability of the inner planet.

\subsection{Effect of General Relativity in Multi-Planet Dynamics}
Previous work has shown that GR can substantially alter the dynamical evolution of multi-planet systems \citep{Adams2006, Faridani2022, Iorio2023}, in some cases suppressing mechanisms such as the Lidov–Kozai resonance \citep{Volpi2024}. GR precession may even be observable in certain configurations \citep{Blanchet2019} and has been found to influence both long-term orbital evolution \citep{Coronel2024} and dynamical stability \citep{Wei2021}.

In this work, we examined models of multi-planet interactions incorporating general relativistic precession and compare them with models in which GR effects are neglected. 
While most planetary systems orbit at distances from their host stars where general relativistic effects are negligible, the habitable zone around a white dwarf lies sufficiently close ($\sim 0.01$ AU) to the stellar surface that such effects cannot be ignored.
 
In the parameter sweep shown in Figure \ref{fig:Csweep}, we explore how the GR apsidal precession modifies the onset of a tidal-driven runaway greenhouse for an inner habitable-zone planet. With GR included (left panel), the stable region (defined as configurations in which the inner planet avoids a runaway greenhouse) extends to significantly smaller values of the semi-major axis of the outer planet and spans a wider range of potential outer planet masses.
This is because GR precession suppresses secularly forced eccentricity growth, limiting tidal heating in the inner planet. 
Without GR (right panel), nearly all configurations with $a_2 \lesssim 0.18$ AU trigger runaway greenhouse conditions, leaving only a small stable region at large separations and low masses. 
In the region of parameter space where GR prevents the onset of a runaway greenhouse, it could act as a protective shield to maintain a planet’s habitable conditions.

If multi-planet systems around white dwarfs are similarly tightly packed to those seen around main-sequence stars \citep{Weiss2018, Weiss2023} or in the pulsar multi-planet population \citep{Wolszczan1992, Wolszczan1994}, then this parameter space is likely of great interest for future habitability and biosignature studies. 
Including GR in modeling of planetary habitability will more accurately predict the level of eccentricity excitation expected for planets near the white dwarf habitable zone and determine whether their atmospheres are likely to be in a runaway greenhouse state.

Although our analysis in this work centers on planets orbiting white dwarfs, similar considerations apply to systems where planets reside at very short orbital radii. Around brown dwarfs or M-dwarfs, general relativistic precession is not significant in the habitable zone, but tidal heating can still influence planetary structure and climate \citep{Seligman2024, Boehm2025}. In such systems, tidal heating might instead be modulated by other sources of precession, such as the stellar quadrupole moment \citep[e.g.,][]{Danby1962, Miralda2002, Li2020}.

\subsection{Future Work}
Planets residing within the habitable zones of white dwarfs remain of considerable interest to the exoplanet community, with ongoing efforts combining JWST observations \citep[e.g.,][]{Kaltenegger_2020, Limbach2022} and theoretical investigations \citep[e.g.,][]{Kozakis2018, Becker2025, Vanderburg2025, Shields2025} to identify targets and evaluate their suitability to maintain habitable conditions. 
The present work has demonstrated the importance of incorporating general relativistic effects into such assessments. 
Several promising avenues for future study remain.

\subsubsection{Planetary Atmosphere Composition}
In this work, we did not examine the detailed chemical composition of the atmosphere; rather, we focused solely on whether a planet could satisfy the \citet{Kasting1988} temperature threshold for triggering a runaway greenhouse effect by using the criterion derived in \citet{BarnesHeller2013}. 
While this approach is sufficient for our present analysis, additional factors are likely to affect the precise limits of the habitable zone. 
For instance, defining habitable zone boundaries for planets orbiting white dwarfs requires detailed climate-state modeling \citep{Zhan2024} as well as consideration of the planet’s landmass fraction \citep{Baker2025}. 
Atmospheric composition will likewise play an important role in refining these boundaries.
Future work assessing the habitability and climate states of planets in the habitable zones of white dwarfs should include both GR precession for the reasons discussed in this work, but also consider the affects of varying atmopshere and surface states.

\subsubsection{Observational Prospects}
One of the top scientific priorities highlighted by the Astro2020 Decadal Survey is to identify and characterize Earth-like planets outside of our solar system, with the long-term goal of imaging these potentially habitable worlds \citep{NAP26141}. 
White dwarfs present unique opportunities for the detection and characterization of potentially habitable planets. 
\citet{Kaltenegger_2020} shows that if an Earth-like planet were discovered in a white dwarf habitable zone, JWST/NIRSpec could reasonably detect key atmospheric biosignatures such as H$_2$O, CO$_2$, O$_2$, O$_3$, CH$_4$, and N$_2$O, were the present only a few transits \citep{Kaltenegger_2020}. 
The mid-infrared capabilities of JWST/MIRI offer a new dimension of constraints to ongoing transit searches \citep{Morris2021}
by detecting thermal emission and IR excesses indicative of planetary atmospheres \citep{Limbach_2022_methods}. 

Given the stability of white dwarf systems over Gyr timescales, coupled with their amenability to detailed spectroscopic characterization, planets in these systems, especially those whose habitability is enhanced by GR suppression of tidal heating, represent some of the most promising near-term candidates for biosignature searches around post-main sequences hosts.





In this work, we have demonstrated that GR can protect against a runaway greenhouse for a single fiducial geometry: a potentially habitable Earth-like ($1 R_{\oplus},\ 1 M_{\oplus}$) planet orbiting at 0.01 AU around a 0.5 $M_{\odot}$ white dwarf with a single external companion planet.
We have shown that for this fiducial parameter set, there is a substantial parameter space where the effects of GR can save a potentially habitably planet from having a runaway greenhouse onset due to perturbations from an additional planetary companion.

For a habitable planet with other parameters (mass, radius, orbital separation), the exact parameter space may look different than that shown in Figure \ref{fig:Csweep}. 
Similarly, additional exterior companions or variations in orbital elements not studied in this work (such as relative inclination) may further affect the exact extent of the parameter space where the runaway greenhouse can be avoided. 
For any specific candidate system discovered in the future, the framework used here can be applied directly by substituting the relevant stellar and planetary parameters into our model to assess whether a companion planet would permit or preclude long-term habitability.


\section{Conclusions} \label{sec:conclusion}
In this work, we have shown that general relativistic apsidal precession can substantially widen the parameter space in which an Earth-like planet in the white dwarf habitable zone can avoid a tidal‐heating–induced runaway greenhouse, despite perturbations from an exterior companion. 
Without GR, nearly all companions within $\sim$0.18 AU trigger uninhabitable conditions for an Earth-like planet orbiting at 0.01 AU around a $0.05 M_{\odot}$ white dwarf. However, including GR suppresses the forced eccentricity and allows many such planets to remain stable and potentially habitable over long timescales. 
While massive or very close companions can still overwhelm this effect, our findings demonstrate that GR can act as a dynamical shield in compact post–main‐sequence planetary systems. 
This protective role should be incorporated into future habitability assessments for planets around white dwarfs and other hosts with similarly compact habitable zones, both to refine theoretical models and to help prioritize observational targets for biosignature searches.



\begin{acknowledgments}
E.S. would like to thank the NASA Wisconsin Space Grant Consortium and the University of Wisconsin-Madison for their generosity in funding undergraduate research.
We would like to thank Dan Tamayo and Andrew Vanderburg for useful conversations. 
We would also like to give special thanks to Sam Hadden for improving the GR capabilities in \celmech.
\end{acknowledgments}

%

\software{Rebound \citep{2012A&A...537A.128R},  
          ReboundX \citep{Tamayo_2019}, 
          Celmech \citep{Hadden_2022}
          }



\bibliography{GRGHE}{}
\bibliographystyle{aasjournalv7}



\appendix

\section{CELMECH  vs. \rebound}
\label{app:nbody}
In this work, we use the secular approximation implemented in \celmech\ \citep{Hadden_2022} instead of performing complete N-body integrations.
Our strategy saves significant computational resources in comparison to using full N-body simulations. 
The secular approximation will be a good model as long as planet semi-major axes are not changing significantly, and dynamics that depend on the planets' instantaneous physical positions (such as mean motion resonances) are not important.
To verify the secular approximation, we ran a test suite of simulations on a coarser grid than our final result in Figure \ref{fig:Csweep} using both \celmech\ \citep{Hadden_2022} and \rebound\ \citep{2012A&A...537A.128R} with GR implemented by \reboundx\ \citep{Nobili1986, Tamayo_2019}. We found no difference in results between the two parameter sweeps, verifying that the secular approach used in this paper is appropriate for our investigations. 

\begin{figure*}[ht!]
\plotone{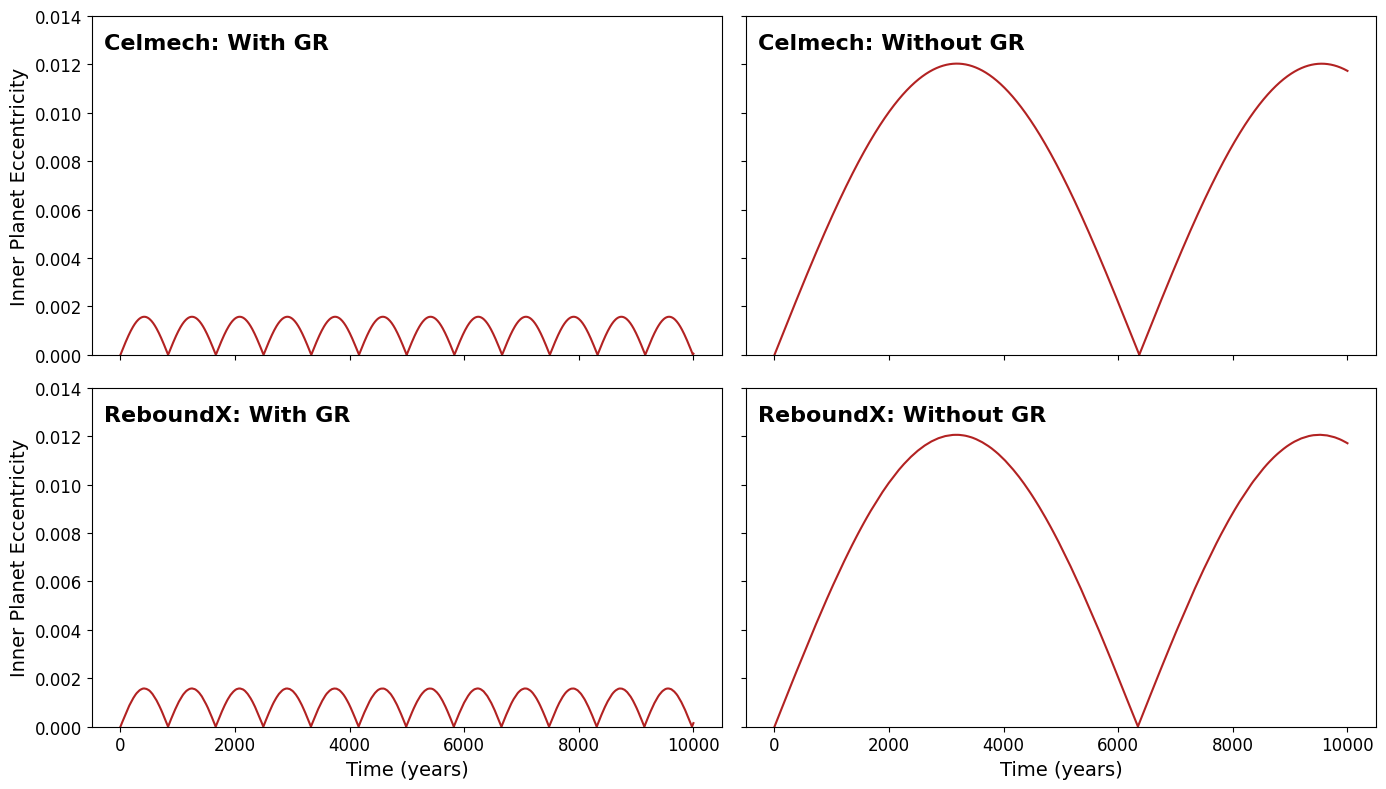}
\caption{Comparison of \reboundx\ and \celmech\ simulations, showing their agreement for integrations run both with and without general relativity. The initial conditions for both the \celmech\ and \rebound\ integrations were identical: an Earth-like inner planet at 0.01 AU with an outer planet of mass 55.36 M$_\oplus$, eccentricity 0.05, and semi-major axis 0.1043 AU.
\label{fig:RvC}}
\end{figure*}

Figure \ref{fig:RvC} compares one example simulation from this grid with the same initial conditions (an Earth-like inner planet with an outer planet of mass 55.36 M$_\oplus$, eccentricity 0.05, and semi-major axis 0.1043 AU) run with \reboundx\ and Celmech for two cases: with and without GR precession. 
As the evolution is secular, there is good agreement between the N-body and secular solutions.

\end{document}